\input harvmac
\parindent 25pt
\overfullrule=0pt
\tolerance=10000

\def\lr { \lref}
\def\de{\Delta}

\def\xxx#1 {{hep-th/#1}}
\def\lr { \lref}
\def\npb#1(#2)#3 { Nucl. Phys. {\bf B#1} (#2) #3 }
\def\rep#1(#2)#3 { Phys. Rept.{\bf #1} (#2) #3 }
\def\plb#1(#2)#3{Phys. Lett. {\bf #1B} (#2) #3}
\def\prl#1(#2)#3{Phys. Rev. Lett.{\bf #1} (#2) #3}
\def\physrev#1(#2)#3{Phys. Rev. {\bf D#1} (#2) #3}
\def\ap#1(#2)#3{Ann. Phys. {\bf #1} (#2) #3}
\def\rmp#1(#2)#3{Rev. Mod. Phys. {\bf #1} (#2) #3}
\def\cmp#1(#2)#3{Comm. Math. Phys. {\bf #1} (#2) #3}
\def\mpl#1(#2)#3{Mod. Phys. Lett. {\bf #1} (#2) #3}
\def\ijmp#1(#2)#3{Int. J. Mod. Phys. {\bf A#1} (#2) #3}
\def\jhep#1(#2)#3{JHEP {\bf #1}(#2)#3}

\lr\grosswitten{D.J. Gross and E. Witten, {\it Superstring modifications of Einstein's equations}, \npb277(1986)1.}
\lr\Muck{W. Muck and  K.S. Viswanathan, {\it Conformal field theory correlators from classical scalar field theory on AdS(d+1)}, \physrev58(1998)041901, \xxx9804035. }
\lr\FreedmanTwo{D.Z. Freedman, S.D. Mathur, A. Matusis and  L. Rastelli, {\it Comments on 4 point functions in the CFT / AdS correspondence}, \xxx9808006.}
\lr\gsw{M.B. Green, J.H.  Schwarz and E. Witten, 'Superstring Theory', (Cambridge University Press, 1986).}
\lr\Maldacena{J. Maldacena, {\it The Large N limit of superconformal field theories and supergravity}, \xxx9711200.}
\lr\GKP{S.S. Gubser, I.R. Klebanov and  A.M. Polyakov, {\it Gauge theory correlators from noncritical string theory}, \plb428(1998)105 \xxx9802109.}
\lr\Witten{E. Witten, {\it Anti-de Sitter space and holography}, \xxx9802150.}
\lr\tHooft{ G. 't Hooft, {\it Dimensional reduction in quantum
gravity}, in 'Salamfest' (World scientific, Singapore, 1993)}
\lr\Susskind{L. Susskind, {\it The World as a hologram}, J.Math.Phys. {\bf 36}(1995)6377, \xxx9409089.}
\lr\SussWitt{L. Susskind and E.  Witten, {\it The Holographic bound in anti-de Sitter space}, \xxx9805114.}
\lr\Nati{S. Lee, S. Minwalla, M. Rangamani and  N. Seiberg, {\it Three point functions of chiral operators in D = 4, N=4 SYM at large N}, \xxx9806074.}
\lr\Gordon{G. Chalmers, H. Nastase, K. Schalm and  R. Siebelink, {\it R current correlators in N=4 superYang-Mills theory from anti-de Sitter supergravity}, \xxx9805105.}
\lr\FreedmanOne{D.Z. Freedman, S.D. Mathur, A. Matusis and  L. Rastelli, {\it Correlation functions in the CFT(d) / AdS(d+1) correspondence}, \xxx9804058.}
\lr\Tseytlin{H. Liu and  A.A. Tseytlin, {\it On four point functions in the CFT / AdS correspondence}, \xxx9807097.}
\lr\bankgreen{T. Banks and M.B. Green, {\it Nonperturbative effects in AdS  x S**5 string theory and d = 4 SUSY Yang-Mills}, \jhep05(1998)002,  \xxx9804170. }
\lr\GreenSethi{M.B. Green and S. Sethi, {\it Supersymmetry constraints on type IIB supergravity}, \xxx9808061. }
\lr\gurarie{V. Gurarie, {\it Logarithmic operators in conformal field theory}, \npb410(1993)535, \xxx9303160.}
\lr\greengut{M.B. Green and M. Gutperle, {\it Effects of D instantons}, \npb498(1997)195, \xxx9701093.}
\lr\HoweWest{P.S. Howe and P.C. West, {\it The complete N=2, d=10  supergravity}, \npb238(1984)181.}
 \lr\Freedmanthree{E. D'Hoker, D.Z. Freedman and W. Skiba, {\it Field
theory tests for correlators in the AdS / CFT correspondence}, \xxx9807098.}
\lr\khoze{N. Dorey, V.V. Khoze, M.P. Mattis and S. Vandoren, {\it
Yang-Mills instantons in the large N limit and the AdS/CFT
correspondence}, \xxx9808157.}
\lr\bianchigreen{M. Bianchi, M.B. Green, S. Kovacs and G. Rossi, {\it Instantons in supersymmetric Yang-Mills and D
instantons in IIB superstring theory}, \xxx9807033.}
\lr\deRoo{M. de Roo, H. Suelmann, A. Wiedemann, {\it The Supersymmetric
effective action of the heterotic string in ten-dimensions},
\npb405(1993)326.}

\Title{\vbox{\baselineskip12pt\hbox{PUPT-1813}\hbox{hep-th/9809067}}}
{\vbox{\centerline{String corrections to four point functions}
       \vskip2pt\centerline{in the AdS/CFT correspondence}}}

\centerline{John H. Brodie
\foot{email:
brodie@puhep1.princeton.edu} 
and Michael Gutperle
\foot{email: gutperle@feynman.princeton.edu}
}
\medskip
\centerline{Department of Physics, Princeton University, Princeton NJ
08554, USA} \bigskip
\centerline{\bf Abstract}
In a string calculation to order $\alpha'^3$, 
we compute an eight-derivative four-dilaton 
term in the type IIB effective action.
Following the AdS prescription, 
we compute the order $(g_{YM}^2N_c)^{-3/2}$ 
correction to the four-point correlation function
involving the operator $tr F^2$ in four dimensional $N=4$ super
Yang-Mills using the string corrected type IIB action 
extending the work of  Freedman et al. (hep-th/9808006). 
In the limit where two of the
Yang-Mills operators approach each other, we find that our correction 
to the four-point
correlation functions developes a logarithmic singularity. We discuss 
the possible cancelation of this logarithmic 
singularities by conjecturing new terms in the 
type IIB effective action.
\medskip
\centerline{\bf PACS: 11.25.Sq,$\;\;$ 11.25.Hf}  
\medskip
\centerline{\bf Keywords: String-theory, AdS-CFT correspondence}
\Date{9/98}

\sequentialequations
\newsec{Introduction}
Despite the fact that our universe is presumed not to have a negative cosmological constant,
there has been a great deal of interest lately in anti-de Sitter gravity.
This is because of the remarkable connection between $d$ dimensional superconformal field theories
decoupled from gravity and $d + 1$ dimensional gravitational theories in the AdS background \Maldacena\GKP\Witten.
Because the degress of freedom of the bulk gravitational theory are described by a field 
theory in one
less dimension living on its boundary, the AdS/CFT correspondence is an example of the
 holographic principle put forward
by \tHooft \Susskind. (see also \SussWitt). In  \GKP \Witten\  it was shown how to
 compute
two-point correlation functions of operators in the conformal field theory by doing a corresponding supergravity
calculation in the AdS background. The prescription given for computing n-point correlation functions is to use the bulk supergravity action as
 the generating function for correlation functions in the boundary conformal field theory
\eqn\op{<O(x_1)...O(x_n)> = {\delta\over \delta\phi_0(x_1)} ... {\delta\over\delta\phi_0(x_n\
)}S^{IIB}_{eff}, }
where $O(x)$ is an operator in the conformal field theory and  $\phi_0$ is the bulk field 
value at the boundary.
Subsequently, three-point correlation functions
were computed in  {\Nati \Gordon \FreedmanOne \Freedmanthree } from the correspondence, while
leading order in $\alpha^\prime$ contributions to four-point correlation functions were examined in  \Tseytlin \FreedmanTwo \Muck.
Unlike two-point and three- point correlation functions which are determined by the conformal
 symmetries,
four-point functions are determined only up to a function of cross ratios \eqn \cross {\eqalign
{\rho & = {x_ {12} x_{34} \over x_{14} x_{23}} \cr \eta & = {x_{12} x_{34} \over x_{13} x_{24}}}.}
In \bankgreen\ it was shown that non-perturbative corrections to four-point-correlation
functions of four dimensional $N = 4$ super Yang-Mills can be calculated in the AdS/CFT 
correspondence
by including stringy $R ^ 4$ corrections to the effective type IIB supergravity action
\eqn\effAction{S^{IIB}_{eff} = { 1 \over \alpha'^4} \int d^{10}x \sqrt {g}
(e^{-2 \phi} R + K \alpha'^3 e^{-\phi
                       \over 2} f_4 (\tau, \bar \tau) R^4)}
since $AdS_5 \times S^5$ is still a solution of \effAction.
The $f_4$ prefactor to the $R^4$ term was argued in \greengut\ to be a generalized Eisenstein
 series
\eqn\fdefi{f_4(\tau,\bar{\tau})=
\sum_{(m,n)\neq(0,0)}{\tau_2^{3/2}\over \mid m\
+n\tau\mid^3}}
which for large $e^{-\phi}$ has the asymptotic expansion
\eqn\expaf{\eqalign{\tau_2^{1/2}f_4&=2\zeta(3)(\tau_2)^{2} +
{2\pi^2\over
3}  + 4\pi^{3/2} \sum_{M}(M\tau_2)^{1/2}\sum_{M|{m}}{1\over
{{m}}^2}
\cr
&
\times  \left(e^{2\pi i M
\tau} + e^{-2\pi i M \bar \tau} \right) \left(1 + \sum_{k=1}^\infty
(4\pi M
\tau_2)^{-k} {\Gamma( k -1/2)\over \Gamma(- k -1/2) k!} \right) ,\cr}}
where $\tau = \chi + ie^{-\phi}$ and $\chi$ is the RR scalar.
The first term in \expaf\ comes from a string tree caluclation, the second from a 1-loop
 calculation, and
the infinite series from charge $M$ D-instantons.
The prefactor \expaf\ has recently been proven by \GreenSethi\ to be the exact coefficient to
the $R^4$ term. Since in the AdS/CFT correspondence  the energy-momentum tensor $T_{\mu\nu}$
of the $d$ dimensional boundary
conformal field theory maps to the graviton in the $d+1$ dimensional
bulk theory,  the four-energy-momentum-tensor
correlation function in four dimesnional $N = 4$ $SU(N_c)$ SYM is
related to the $R^4$ term of IIB superstring theory via \op after substituting \effAction.
The prefactor \expaf\ expressed in terms of   Yang-Mills\
 parameters,
$\lambda =  g^2_{YM} N_c = {L^4 \over \alpha'^2}$ and $ g^2_{YM} = 4 \pi
g_s$ takes the following form
\eqn\em{ f_4(\lambda,N_c)=
{2\zeta(3)\over 4\pi\lambda^{3/2}} + {\pi^2\lambda^{1/2}\over 2N_c^2(4\pi)^{1/2}}
 + {(4\pi)^{3/2}\over N_c^{3/2}} \sum_M Z_M M^{1/2} (e^{-M({8\pi^2\over g^2_{YM}} + i\theta)}
+  e^{-M({8\pi^2\over g^2_{YM}} - i\theta)}).} 
Note that the last term in \em\ comes from Yang-Mills instantons  and
this indicates that D-instantons in $AdS_5$ can be identified with
instantons in the SYM \bianchigreen,\khoze.

Interestingly  there are many other terms in the IIB effective action that are related to the $R^4$ term
by supersymmetry. We will show that
one such term involving only the scalar dilaton field is
\eqn \scalar {{1\over \alpha'^4 } \int d^{10}x \sqrt {-g }
\alpha'^3 e^{- \phi/2 } f_4 ( \tau, \bar \tau ) D_{\mu} D_{\nu}\phi(x)
D^{\mu}D^{\nu}\phi(x)
D_{\alpha} D_{\beta} \phi(x)D^{\alpha}D^{\beta}\phi(x).}
where $\phi(x)$ is the dilaton.
In this note we calculate the $(g_{YM}^2N_c)^{-3/2}$ correction  to the four-point correlation function of the trace of the
$N=4$ Yang-Mills field strength squared, $tr (F^2)$,
from the term \scalar\ in the IIB effective action expanded about the
$AdS_5 \times S^5$ background.
In section 2, we derive  \scalar\ by doing a string theory 
calculation in Minkowski space. In section 3, we compactify  IIB string theory on
$AdS_5\times S_5$ and in the Maldacena limit \Maldacena\ we calculate the result for
the four point correlation function of $tr(F^2)$ using the 
AdS/CFT correspondence. In section 4, we show 
that in the limit where two operators in the CFT  approach each other 
the contribution to the four-point  functions calculated in section 3
exhibit  logarithmic singularities.
If these logarithmic singularities were  really present this would
have  rather dramatic consequences for the conformal field theory;
namely, it would imply  that $N=4$ SYM is a logarithmic CFT \gurarie. In
section 5, we discuss how including terms in the effective action containing the Riemann
curvature and five form field strenth and four scalars 
could cancel the logarithmic singularities.

\newsec{Four-point functions in Minkowski space}
The tree four-point scattering amplitudes for the
massless states of type IIB superstring theory can be expressed in a compact form
\eqn\fourpoint{A=c{\cal A}_{tree}(s,t,u)
K(\zeta_1,k_1,\cdots,\zeta_4,k_4)K(\bar{\zeta}_1,k_1,\cdots,\bar{\zeta}_4)}
where the tree level amplitude is given by
\eqn\tree{A_{tree}={1\over
g_{st}^2}{\Gamma(-s/4)\Gamma(-t/4)\Gamma(-u/4)\over 
\Gamma(1+s/4)\Gamma(1+t/4)\Gamma(1+u/4)}.  }
The kinematic factor $K\bar{K}$ is the same for the tree and one loop
amplitude and is the product of two kinematic factor for the  open string four
point function. This reflects the fact that the closed string can be viewed as
the product of a left and rightmoving open string. In  the kinematic
factor $K$ for the open string the wavefunction
$\zeta_i$ for the $i$-th particle  can either be a
vector $\xi_\mu$ or a Majorana-Weyl spinor $u_a$.
For four vectors the kinematic factor takes the form
\eqn\fourvect{\eqalign{K(\xi_1,\xi_2,\xi_3,\xi_4)&=
t^8_{\mu\nu\lambda\kappa\rho\sigma\tau\omega} \xi_1^\mu k_1^\nu
\xi_2^\lambda
 k_2^\kappa
\xi_3^\rho k_3^\sigma \xi_4^\tau k_4^\omega\cr
&= -{1\over 4}\big(st\xi_1\cdot\xi_3 \xi_2 \cdot \xi_4+su\xi_2\cdot
\xi_3\xi_1 \cdot \xi_4 +\cdots\big)}}
Where the detailed expression for $t_8$ can be found in \gsw. 
The massless states of the IIB superstring are given by the tensor
product $\zeta_L\otimes\zeta_R$. The bosonic states are given by
$\xi_\mu\otimes\xi_\nu= h_{(\mu\nu)}+b_{[\mu\nu]}$ where $b$ denotes the NS-NS antisymmetric
tensor and $h$ contains the transverse and traceless graviton
together with the transverse \
dilaton.
In the following we will be particularly interested in the four point
function for the dilaton $\phi$.

The scattering amplitude \fourpoint\ contains an infinite number of
poles in the s,t and u channel, coming from massive intermediate string
states. In a low energy effective field approximation the effect of this stringy
behavior can be seen by an $\alpha^\prime$ expansion of \fourpoint\
which effectively integrates out all the massive intermediate string
states at tree-level and  leads to an infinite\
 series of higher derivative contact terms. This expansion was
performed in \grosswitten\
\eqn\grw{{\Gamma(-1-s/8)\Gamma(-1-t/8)\Gamma(-1-u/8)\over
\Gamma(2+s/8)\Gamma(2+t/8)\Gamma(2+u/8)}= {1\over stu}+2\alpha'^3\zeta(3)+O(\alpha^{\prime\; 5})}
Since the kinematic tensor in \fourpoint\ contain eight momenta, it is easy to see 
on dimensional grounds that 
the first term in \grw\  
corresponds to a two derivative contact term and does not give  a new
term in the effective action. This is produced
by the lowest order terms in the action like the Einstein term. 
The next term in \grw\ 
corresponds to a new eight-derivative string theory correction 
 in  the IIB supergravity effective action.
We will focus on the eight derivative four dilaton amplitude which will be related to
the four point function of $\tr(F^2)$ in the AdS/CFT correspondence.

The dilaton polarization tensor is given by $\xi_\mu\otimes
\xi_\nu=\phi/\sqrt{8}\big(\eta_{\
\mu\nu}-n_\mu k_\nu -n_\nu k_\mu)$. The $n_\mu$
satisfy $n^2=\
0,n\cdot k=1$ and are neccessary to make the dilaton polarization
vector transverse, i.e. $\zeta_{\mu\nu}k^\nu=0$. Plugging this into
$K\bar{K}$ with $K$ and \
$\bar{K}$ given by \fourvect\ leads to
\eqn\foudila{\eqalign{K\bar{K}&=t_8^{\alpha_1\alpha_2\alpha_3\alpha_4\alpha_5
\alpha_6\alpha_7\alpha_8}t_8^{\beta_1
\beta_2\beta_3\beta_4\beta_5\beta_6\beta_7\beta_8 }k^1_{\alpha_1}k^1_{\beta_1}\xi^1_{\alpha_2\beta_2}\cdots
k^4_{\alpha_7}\
k^4_{\beta_7}\xi^4_{\alpha_8\beta_8}
\cr&={\phi^4\over 16} \big(s^2t^2+s^2u^2+t^2u^2\big)={\phi^4\over 32} \big(s^4+t^4+u^4)}}
the normalization $\phi$ of the dilaton vertex can be fixed by
comparing the three point functions involving two dilatons and one
graviton with the three point function coming from the kinetic term
for the dilaton in the IIB effective action.
The eight derivative term which reproduces this S-matrix element is
given by
\eqn\fourdil{{ S}^{(1)}_{\phi^4}=c {\zeta(3)\alpha^{\prime 3}\over g_s^2}\int
d^{10}x \sqrt{g}D_{\mu}D\
_{\nu}\phi
 D^{\mu}D^{\nu}\phi D_{\rho}D_{\lambda}\phi D^{\rho}D_{\lambda}\phi}

Apart from the tree level contribution \tree, the four point amplitudes 
at order $\alpha'^3$ receives a one-loop and non-perturbative D-instanton contributions. 
The exact form was conjectured in \greengut\ to be
\eqn\fourptsl{A_4=K\bar{K} \tau_2^{1/2}f_4(\tau,\bar{\tau})}
where $f_4$  \fdefi\ is a nonholomorphic  Eisenstein series.
In the weak coupling expansion \expaf\ of \fourptsl\ the first two term are the tree level and one loop contributions,
whereas the rest is an infinite series of D-instanton contributions and
perturbative fluctuations around the D-instantons.

The form of the terms in the IIB effective action which obey a non renormalization
theorem \GreenSethi\ can be encoded in an integral over half the IIB superspace
\HoweWest\ of
an appropriate power of a a constrained linearized 
superfield $\Phi(x,\theta)$ where $\theta^a,a=1,\cdots, 16$. The superfield
satisfies
 \eqn\constr{\bar{D}\Phi=0,\quad\quad D^4 \Phi
=\bar{D}^4\bar{\Phi}}
and has an expansion up to order $\theta^8$ given by
\eqn\superf{\Phi= \tau + \bar{\theta}\lambda+ \cdots \theta^8
D^4\bar{\tau}}
where the complex scalar $\tau$ contains the dilaton and RR-scalar
$\tau=\chi+i\
e^{-\phi}$.
The four point function can the  be expressed as an integral over half
the superspace
\eqn\intsupp{\eqalign{{Re}\Big(\int d^{16}\theta \Phi^4\Big) = {1\over
2}\Big(\tau^2D^4\bar{\tau} D^4 \bar{\tau}+\bar{\tau}^2D^4{\tau} D^4
{\tau}\Big)\cr
=D_\mu D_\nu \tau D^\mu D^\nu \bar{\tau}D_\rho D_\lambda \tau D^\rho D^\lambda \bar{\tau}}}
where we integrated by parts.

\newsec{Four point function in AdS/CFT}
The AdS/CFT correspondence allows one to calculate correlation functions
of operators in the $N=4$ SYM using the effective action of IIB
superstring theory. We will focus on terms in the effective action
which involve eight derivatives on  four dilatons. The dilaton $\phi$ of IIB
on $AdS_5\times S_5$ corresponds to the marginal operator $\tr(F^2)$.
 
We will concentrate on an $\alpha^{\prime\; 3}$  four dilaton term
\fourdil\ compactified on $AdS_5\times S_5$  which is given by
\eqn\fourdilb{{ S^{(1)}}_{\phi^4}=\int {d^5z\over z_0^5} \sqrt{g}D_{\mu}D_{\nu}\phi
 D^{\mu}D^{\nu}\phi D_{\rho}D_{\lambda}\phi D^{\rho}D_{\lambda}\phi.}
 where we drop the $S_5$ dependence (which only
gives a volume factor). 
The prescription to calculate CFT correlators using the IIB efective
action is given by 
\eqn\precr{W_{CFT}(\phi_{CFT})= S_{IIB}(\phi_{bulk})}
where $W_{CFT}$ in the generating function for the conformal field theory.
Here $S_{IIB}$ is the IIB effective action compactified on
$AdS_5\times S_5$ and the bulk and CFT fields are related (for scalars
of dimension $\de$) by
\eqn\relsc{\phi_{bulk} (z) = c \int {d^4x}K_{\de}(z,
x)\phi_{CFT}(x)    }
A correlator of operators in the CFT associated with $\phi_{CFT}$ can
then be calculated by
\eqn\corrcalc{\eqalign{I(x_1,x_2,x_3,x_4)&= \langle {\cal O}(x_1){\cal
O}(x_2){\cal O}(x_3){\cal O}(x_4)\rangle_{CFT}\cr
&={\delta\over \delta \phi(x_1)}{\delta\over \delta
\phi(x_2)}{\delta\over \delta \phi(x_3)}{\delta\over \delta \phi(x_4)}W(\phi)} }
In order to calculate the correction to the correlator $I=\langle
\tr(F^2)\tr(F^2)\tr(F^2)\tr(F^2)\rangle$ coming from the higher
derivative term \fourdilb, and use the prescription \corrcalc\ which gives
\eqn\corrs{I^s=\int {d^4z dz_0\over z_0^5}D_\mu D_\nu K_{\de}(z,x_1)
D^\mu D^\nu K_{\de}(z,x_2)D_\rho D_\lambda K_{\de}(z,x_3)
D^\rho D^\lambda K_{\de}(z,x_4)}
Here $I^s$ denotes the 's-channel' part of the amplitude. There are
also $I^t$ and $I^u$ which are obtained from \corrs\ by exchanging
$x_2 \leftrightarrow x_4$ and $x_2 \leftrightarrow  x_3$ respectively, and the complete correlator is given
by $I=I^s+I^t+I^u$.

The metric on the five dimensional  euclidean AdS space is given by
\eqn\metr{g_{\mu\nu}={1\over z_0^2}\delta_{\mu\nu}, \quad\quad \mu,\nu=0,\cdots\
,4}
Hence the nonvanishing components of the metric connection are given by
\eqn\connec{\Gamma^0_{ij}={1\over z_0}\delta_{ij}
,\quad \Gamma^i_{0j}=-{1\over \
z_0} \delta_{ij},\quad \Gamma^0_{00}=-{1\over z_0} }
where $i,j = 1...4$.
The bulk to boundary Greens function for a field of conformal dimension $\Delta$
is given by
\eqn\prop{K_\Delta=\lambda_\Delta \left({z_0\over z_0^2+(z-x)^2}\right)^\Delta}
For notational simplicity we leave out the normalization factor
\eqn\norm{\lambda_\de =\Gamma(\Delta)/\big(\pi^{d/2}\Gamma(\Delta-d/2)\big)}
from \prop\  in the following  which can be easily
reinstated at the end.
The second order covariant 
derivatives with respect to the connection \connec\ of the propagtor $D_\mu
 D_\nu K_\Delta(z,x)$ are given by
\eqn\secder{\eqalign{D_0D_0 K_\Delta &= \Delta^2{z_0^{\Delta-2}\over
(z_0^2+(z-x)^2)^\Delta} -4\Delta(\Delta+1){z_0^{\Delta}\over
(z_0^2+(z-x)^2)^{\Delta+1}}\cr
&+4\Delta(\Delta+1){z_0^{\Delta+2}\over
(z_0^2+(z-x)^2)^{\Delta+2}},\cr
D_0D_i K_\Delta &= -2\Delta(\Delta+1){z_0^{\Delta- 1}\over
(z_0^2+(z-x)^2)^{\Delta+1}}(z-x)^i\cr&
 -2\Delta(\Delta+1){z_0^{\Delta+1}\over
(z_0^2+(z-x)^2)^{\Delta+2}}(z-x)^i,\cr
D_iD_j K_\Delta &= -\Delta\delta_{ij}{z_0^{\Delta- 2}\over
(z_0^2+(z-x)^2)^{\Delta}}
+4\Delta(\Delta+1){z_0^{\Delta}\over
(z_0^2+(z-x)^2)^{\Delta+2}}(z-x)^i(z-x)^j
.} }
To calculate $g^{\mu\nu}g^{\rho\lambda}D_\mu D_\rho K_\Delta(z
,x_i)D_\nu D_\lambda K_\Delta(z,x_j)$ we use the formulas \secder\ and
the metric
\metr. 
The resulting expression can be simplyfied using
$(z-x_i)\cdot(z-x_j)=1/2\{(z-x_i)^2+(z-x_j)^2-(x_i-x_j)^2\}$.  The
result is given by the identity
\eqn\contract{\eqalign{g^{\mu\nu}g^{\rho\lambda}D_\mu D_\rho K_\Delta(z
,x_i)D_\nu D_\lambda K_\Delta(z,x_j)=
\Delta^2(\Delta^2+d)K_\Delta(z,x_i)K_\Delta(z,x_j)\cr
-4\Delta^2(\Delta+1)^2x_{ij}^2
K_{\Delta+1}(z,x_i)K_{\Delta+1}(z,x_j)+4\Delta^2(\Delta+1)^2x_{ij}^4
K_{\Delta+2}(z,x_i)K_{\Delta+2}(z,x_j)}}
Note that once \contract\ is expressed in terms of $K_\de$'s, all
terms with extra factors of $z_0$ cancel. This fact will be important for the
conformal invariance of the resulting CFT correlators.

Using \contract\  $I^s$ can be brought into the following form 
\eqn\resfour{\eqalign{I^s= \de^4(\de^2+d)^2
I_{\de\de\de\de}-4\de^4(\de+1)^2(\de^2+d)\Big\{x_{12}^2
I_{\de+1\de+1\de\de}+x_{34}^2 I_{\de\de\de+1\de+1}\cr-x_{12}^4
I_{\de+2\de+2\de\de} -x_{34}^4 I_{\de\de\de+2\de+2}\Big\}
+16\de^4(\de+1)^4\Big\{x_{12}^4x_{34}^4
I_{\de+2\de+2\de+2\de+2}\cr-x_{12}^2x_{34}^4
I_{\de+1\de+1\de+2\de+2}-x_{12}^4x_{34}^2
I_{\de+2\de+2\de+1\de+1}+x_{12}^2x_{34}^2
I_{\de+1\de+1\de+1\de+1}\Big\}}}
 where  $I_{\de_1\de_2\de_3\de_4}$ is defined  by
\eqn\Idefin{I_{\de_1\de_2\de_3\de_4}= \int {d^5z\over
z_0^5}K_{\de_1}(z,x_1)K_{\de_2}(z,x_2)K_{\de_3}(z,x_3)K_{\de_4}(z,x_4)}
$I^t$ can be obtained from  \resfour\ by permuting $x_2\leftrightarrow x_4$ and the
arguments $\de_2\leftrightarrow \de_4$ in  $I_{\de_1\de_2\de_3\de_4}$
and  similarly  $I^u$ can be obtained  by  $x_2\leftrightarrow x_3$ and the
arguments $\de_2\leftrightarrow \de_3$ in  $I_{\de_1\de_2\de_3\de_4}$.

\newsec{Logarithmic singularities in the four point function}
We want to investigate the  corrections to correlators of the four dimensional boundary conformal field theory 
coming
from the eight derivative terms $I^s,I^t$ and $I^u$ of the bulk supergravity theory. 
In particular we
want to analyze the behavior of these four point functions when two
operators come close, i.e, $x_{ij}\to 0$
with all the other separations remaining finite. 
The most useful representation of $I_{\Delta_1\Delta_2\Delta_3\Delta_4}$ for this purpose is given by (51) in the appendix
\eqn\Iintrep{I_{\Delta_1\Delta_2\Delta_3\Delta_4}= C \int
d\beta_2d \beta_4
{\beta_2^{\de_2-1}\beta_4^{\de_4-1}\over \big(
x_{13}^2+\beta_2x_{23}^2+\beta_4x_{34}^2\big)^{\de_3}\big(\beta_2x_{12}^2+\beta_4x_{14}^2+\beta_2\beta_4x_{24}^2\big)^{\de/2-\de_3}}}
Here we denote $\de=\sum_{i=1}^4\de_i$.
It is easy to see that if $x_{12}\to 0$ and all other $x_{ij}$ remain
finite the only region of the integration which can give a singularity
is $\beta_4\to 0$. Extracting $\beta_2$ from the second term in the
denominator and redefining $\beta_4^\prime= \beta_4
(x_{14}^2+\beta_2x^2_{24})/\beta_2$ gives
 \eqn\logsina{\eqalign{I_{\Delta_1\Delta_2\Delta_3\Delta_4}= C \int d\beta_2d\beta_4
{d\beta_4^{\de_4-1}\over
(x_{12}^2+\beta_4)^{\de/2-\de_3}}{\beta_2^{\de_2+\de_3+\de_4-\de/2-1}\over
\big(x_{14}^2 +\beta_2x_{24}^2\big)^{\de_4}}\cr
\times{1\over 
\big(x_{13}^2+\beta_2x_{23}^2+
\beta_4\beta_2x_{34}^2/(x_{14}^2+\beta_2x_{24}^2)\big)^{\de_3}
}}}
It is easy to see that \logsina\ displays a divergence
$1/(x_{12}^2)^k$ as $x_{12}^2\to 0$ if $\de_3+\de_4=\de/2+k$ and it
displays a logarithmic divergence $\ln(x_{12}^2/x_{13}x_{14})$ if
$\de_3+\de_4=\de/2$. Note that if there is a power singularity of
certain order  in \logsina\ then expanding the rest of the integral as
a power series in $\beta_4$ also produces lower order singularities.

As we shall see only the logarithmic singularities will be important
in the calculation of the four point function. Setting $\de_3+\de_4=\de/2$,
 the singular part of the integral\foot{We drop nonsingular terms which are important for conformal invariance.} then becomes
\eqn\singint{I_{\Delta_1\Delta_2\Delta_3\Delta_4}\sim C {1\over
(x_{13}^2)^{\de_3}}{1\over (x_{14}^2)^{\de_4}} \ln\left({x_{12}^2\over x_{13}x_{14}}\right)\int d\beta_2
{\beta_2^{\de_2-1}\over (1+\beta_2)^{\de/2}}+o(1)}  
In this limit the integral over $\beta_2$ simply gives a Euler beta
function and together with the value of $C$ given in (52) the  coefficient
of the log singularity is given by
\eqn\logsing{I_{\Delta_1\Delta_2\Delta_3\Delta_4}\sim {\pi^2\over 2}{1\over
(\de/2-1)(\de/2-2)}{1\over
(x_{13}^2)^{\de_3}}{1\over (x_{14}^2)^{\de_4}}\ln\left({x_{12}^2\over x_{13}x_{14}}\right)+o(1)}

The behavior of $I^s$ in \resfour\  as $x_{12} \rightarrow 0$ is easy to analyze,
all $I_{\Delta_1\Delta_2\Delta_3\Delta_4}$
which produce power singularities, i.e. have $\de_4+\de_4>\de/2$ are
multiplied by factors of $x_{12}^2$ such that the result is
nonsingular. In addition the only logarithmic singularity comes from
the first term in \resfour\ since all the other logarithmic divergent
terms are also multiplied by factors of  $x_{12}^2$.

On the other hand one finds that all the terms in $I^t$ and $I^u$ display logarithmic
singularities as $x_{12}^2\to 0$,  and the complete logarithmic singularity is given by
summing over all terms taking the prefactors appearing in \resfour\
and \logsing\ into account. The final result is
\eqn\finallog{I\sim  c_1  {1\over (x_{13}^2)^{\de_3}}{1\over (x_{14}^2)^{\de_4}}\ln(x_{12}^2/x_{13}x_{14})}
where the numerical constant $c_1$ is given by
\eqn\cone{c_1= {\pi^2 \lambda_\de^4\over 2}\Big\{{3\de^4(\de^2+d)^2\over (2\de-1)(2\de
-2)}-{32\de^4(\de+1)^2(\de^2+d)\over (2\de+1)(2\de)(2\de-1)}+ {16\cdot
12 \de^4(\de+1)^4\over (2\de+3)(2\de+2)(2\de+1)2\de}\Big\}}
where we reinstated the normalization $\lambda_\de$ defined in \norm\
of the propagators \prop.
It is easy to see that $c_1$ does not vanish, in particular for the
case of the dilaton and $AdS_5$ we set $d=4$ and $\de =4$ and get
\eqn\insfour{c_1= {1\over \pi^6} {152985600\over 77}}

\newsec{Possible cancellations of logarithmic terms}
In the last section we found we found a logarithmic singularity in the
four point function of $N=4$ SYM in the large $N_c$ limit at order
$1/(g_{YM}^2N_c)^{3/2}$. Similar logarithmic singularities at leading order 
in four-point functions were found
in \FreedmanTwo\ coming from the tree level two derivative part of the
action. In \FreedmanTwo, it was speculated that by adding all diagrams the
logarithmic singularities might cancel. We will now adress this possibility of cancelation of 
logarithmic singularities in our case. We found a nonzero logarithmic
contribution from summing the $s,t$ and $u$ channel of diagrams coming
from \fourdil. Hence additional four point functions which might cancel the
logarithmic singularities in \finallog\ must come from other terms in the effective action.\foot{We are grateful to A. Rajaraman for raising this possibility.}
This is possible because in contrast to the flat Minkowski background
the $AdS_5$ background has a nonvanishing curvature tensor as well as
a nonvanishing five-form field strength.
\eqn\curvads{\eqalign{R_{\mu\nu\lambda\rho}&=-{1\over
L^2}\big(g_{\mu\lambda}g_{\nu\rho}-g_{\mu\rho}g_{\nu\lambda}\big)\cr
F_{\mu\nu\lambda\rho\sigma}&= {1\over L}\epsilon_{\mu\nu\lambda\rho\sigma}}}
Hence there is the possibility that terms of the same order in $\alpha'$ as the eight derivative term
\fourdil\ which involve  curvature tensors or five form field
strengths together with four scalars  contribute  to the  four point
function in the conformal field theory once the constant background
values  \curvads\ are 
substituted.  Examples of such terms are four dilatons with six
derivatives together with one Riemann ternsor or two $F_5$ as well as 
terms with two curvature tensors,four $F_5$ or one Riemann tensor and
two $F_5$ and four dilatons
with four derivatives. Such terms should be in the supersymmetric completion of the
$R^4$ term like the eight derivative term of four scalars.
After using \curvads\ the generic form of the additional four-scalar terms in the
effective action is given by
\eqn\addterms{\eqalign{S_{\phi^4}^{(2)}&\sim\int {d^5 z\over z_0^5} D_\mu \phi D^\mu \phi
D_\nu D_\rho \phi D^\nu D^\rho \phi \cr
S_{\phi^4}^{(3)}&\sim \int {d^5 z\over z_0^5} D_\mu \phi D^\mu \phi
D_\nu \phi D^\nu  \phi}.}
Note that there is an additional  the eight derivative
term where the indices are contracted cyclicly. 
In flat space this  contraction of the derivatives is
 related to \fourdil\ by integration by parts. In the
AdS background the issue of the noncommutative nature of the covariant
derivatives becomes important and an integration by parts relates the
two eight derivative terms  up to terms involving the curvature and two less derivatives
on the scalars, since
\eqn\ThreeDer{[D_{\mu},D_{\nu}]D_{\rho}\phi = R^{\lambda}_{\mu\nu\rho}D_{\lambda}\phi.}

It is easy to repeat the analysis of section 4  for the terms
 \addterms\ of the effective action and it follows that the 
logarithmic singularity of the four point correlation function of
the CFT
as $x_{12}^2\to 0$ induced by these terms in exactly of the same form as \finallog.
The result for the numerical coefficients $c_2$ and $c_3$ multiplying
the singularities is 

\eqn\resultaddterms{\eqalign{c_2&={\pi^2 \lambda_\de^4\over 2}\Big\{ {(8-2\de)(\de^6+\de^4d)\over (2\de-1)(2\de-2)2\de}+{32\de^4(\de+1)^2(\de-2)\over (2\de+2)(2\de+1)2\de(2\de-1)}\Big\} \cr
c_3&={\pi^2\lambda_\de^4\over 2}\Big\{  {3\de^4\over (2\de-1)(2\de-2)}-{8\de^4\over
2\de(2\de-1)}+{8\de^4\over (2\de+1)2\de}\Big\}}}

For the four dilaton amplitude in  $AdS_5$ inserting $\de=4,d=4$ into
\resultaddterms\ gives  

\eqn\muresadd{c_2= {1\over \pi^6}{ 368640\over 7},\quad c_3=
{1\over \pi^6}{46080\over 7}}
The supersymmetric completetion of the $R^4$ term in the ten dimensional IIB effective action 
should in principle
uniquely determine the relative coefficents of these
terms. Unfortunately the details of this are not known and either
the supersymmetrization of the $R^4$ term 
or a calculation of five point and six point
S-matrix elements in string perturbation 
theory involving one or two gravitons and
four dilatons around flat space 
to determine the relative normalizations of the three terms 
seems to be  difficult\foot{\deRoo\ for a  discussion of the  
supersymmetrizations of $R^4$ in the $d=10$, N=1 context.}. We are not able to confirm or defute the
cancellation of the logarithmic singularities coming from the $\alpha'^3$
terms in the IIB effective action at this moment. 
\newsec{Conclusions}
In this note, we computed the corrections to the four-point
correlation funcetion of the operator $trF^2$ from the
eight-derivative-four-dilaton term in the IIB effective action using
the AdS/CFT correspondence. We found that as two of the Yang-Mills
operators approach each other, the four point function yields a
logarithmic singularity, like the logarithmic singularities found at a
lower order in \FreedmanTwo. This may indicate that the theory is a
logarithmic conformal field theory. Logarithmic conformal field
theories in two-dimensional are non-unitary. If this were the case in
four-dimensions, it would certainly be a shocking result. It would
imply that $AdS$ gravity is described holographically by a non-unitary
conformal field theory!  Clearly, the most satisfying solution would
be if the logarithmic singularities in our four-point calculation
\finallog\ were 
cancelled by other terms in the effective action. However, the
coefficients required for cancellation seem perculiar and
unfortunately are difficult to calculate. However, the fact that the
logarithmic singularities do appear in our calculations and in
\FreedmanTwo\ seems to suggest that they are a generic feature of the
AdS/CFT correspondence. At present there is the possibility  
that the CFT on the boundary of AdS either maybe logarithmic  if a four dimensional
 analog of \gurarie\ exists or that a new scale is introduced 
and the theory ceases to be conformal. 

Assuming that the logarithic singularities do cancel, the $I_{s,t,u}$
terms will modify the CFT four-point function with finite terms which
is in agreement with the intuition that the massive string modes which
when integrated out do not modify the structure of the CFT significantly.

\bigskip
\noindent{\bf Acknowledgments}
\smallskip
M.G. gratefully acknowledges the hospitality of the Theory divison
at CERN at early stages of this work. J.H.B and M.G.  are 
grateful to the Aspen Center of Physics for providing a stimulating
reserach environment while this work was completed.  Its a pleasure to thank 
G. Chalmers, D. Freedman, E. Gimon, M.B. Green, V. Gurarie,  A. Lawrence, H. Liu, 
S. Mathur and especially A. Rajaraman for useful conversations and comments.
The work of M.G. is partially supported by DOE grant
DE-FG02-91ER40671 and NSF grant PHY-9157482.
\bigskip
\noindent{\bf Appendix} 
\bigskip
The integrals which appear in the calculation of the four point amplitudes can \
be partially done  using the method of Feynman parameters. The basic formula is
\eqn\fmpara{{1\over X_1\cdots X_n}= \int d\alpha_1\cdots d\alpha_n \delta(1-\sum\alpha_i){1\over  \big( \alpha_1X_1+\cdots \alpha_nX_n\big)^n}}
Specializing to $n=4$ and taking derivatives with respect to  $X_i$ the following 
formula easily follows
\eqn\fmptwo{{1\over X_1^{\de_1}\cdots X_4^{\de_4}}= {(\de-1)!\over 6
(\de_1-1)!\cdots ( 
\de_4-1)!}\int d\alpha_1\cdots d\alpha_4 \delta(1-\sum\alpha_i)  
{\alpha_1^{\de_1-1}\cdots \alpha_4^{\de_4-1}\over 
\big( \alpha_1X_1+\cdots+ \alpha_4X_4\big)^\de}}
Where $\de=\sum_{i=1}^4 \de_i$. In the supergravity amplitudes
 the factors $X_i$ are given by $X_i=z_0^2+(z-x_i)^2$. It is easy to see that
\eqn\denoone{\alpha_1X_1+\cdots \alpha_4X_4= z_0^2+z^2+
\sum_{i<j}\alpha_i\alpha_j x_{ij}^2}
where $x_{ij}^2=(x_i-x_j)^2$. The introduction of Feynman parameters makes it straightforward to do the integral over the $AdS_5$ in the four point function. The general form of such integrals is
\eqn\adsint{\eqalign{I(N,M)&=\int {dz_0 d^4z\over z_0^5} {z_0^{2N+1}\over \big(z_0^2+z^2+\sum_{i<j}\alpha_i\alpha_j x_{ij}^2\big)^M}\cr
&={\pi^2\over2} {N!(M-N-4)!\over (M-1)!} {1\over\big(\sum_{i<j}\alpha_i\alpha_j\
 x_{ij}^2\big)^{M-N-3}}}}
 The integrals which we want to evaluate 
in the AdS bulk are in general of the following form
\eqn\inteval{I_{\Delta_1\Delta_2\Delta_3\Delta_4}=\int{dz_0\;d^4z\over z_0^5}  K_{\Delta_1}(z,x_1)K_{\Delta_2}(z,x_2)K_{\Delta_3}(z,x_3)K_{\Delta_4}(z,x_4)}
where $K_\Delta$ is defined in \prop. Using \fmptwo\ gives
\eqn\inttwo{\eqalign{I_{\Delta_1\Delta_2\Delta_3\Delta_4}&={(\Delta-1)!\over 6 (\Delta_1-1)!\cdots (\Delta_4-1)!}\int d\alpha_1\cdots d\alpha_4 \delta(1-\sum\alpha_i)\cr
&\times\int{ dz_0\;d^4z\over z_0^5}
 z^{\Delta}{\alpha_1^{\Delta_1-1}\cdots \alpha_4^{\Delta_4-1}\over \big(  z_0^2+z^2+\sum_{i<j}\alpha_i\alpha_j x_{ij}^2)^\Delta}}}
Hence the integral over the AdS bulk variable $z$ can be done using
\adsint\  giving
\eqn\intthree{\eqalign{I_{\Delta_1\Delta_2\Delta_3\Delta_4}&={\pi^2\over2} \
{(\Delta/2-3)!(\Delta/2-1)!\over (\Delta_1-1)!(\Delta_2-1)!(\Delta_3-1)\
!(\Delta_4-1)!}\cr
&\times\int d\alpha_1\cdots d\alpha_4 \delta(1-\sum\alpha_i){\alpha_1^{\Delta_1-1}\cdots \alpha_4^{\Delta_4-1}\over \big(\sum_{i<j}\alpha_i\alpha_j x_{ij}^2)^{\Delta/2}} }}
In \Muck\ such a integrals were evaluated by introducing new variables
$\beta_1=
\alpha_1,\beta_2=\alpha_1\alpha_2,\beta_2=\alpha_1\alpha_3,\beta_4=
\alpha_1\alpha_4$.The integral over $\beta_1$ can than be done by using the $\delta$-function constraint. 
\eqn\intfour{\eqalign{&\int d\alpha_1\cdots d\alpha_4
\delta(1-\sum\alpha_i){\alpha_1^{\Delta_1-1}\cdots
 \alpha_4^{\Delta_4-1}\over \big(\sum_{i<j}\alpha_i\alpha_j
x_{ij}^2)^{\Delta/2}}\cr
=&\int d\beta_2 d\beta_3 d\beta_4
 {\beta_2^{\Delta_2-1}\beta_3^{\Delta_3-1}\beta_4^{\Delta_4-1}\over \big(\
\beta_2 x_{12}^2+\beta_3 x_{13}^2+\beta_4 x_{14}^2+\beta_2\beta_3
x_{23}^2+\beta_2\beta_4 x_{24}^2+\beta_3\beta_4
x_{34}^2\big)^{\Delta/2}}  \cr
=& {\de_3-1)!(\de/2-\de_3-1)!\over (\de/2-1)!}\int d\beta_2d \beta_4
{\beta_2^{\de_2-1}\beta_4^{\de_4-1}\over \big(
x_{13}^2+\beta_2x_{23}^2+\beta_4x_{34}^2\big)^{\de_3}}\cr
\times& { 1\over \big(\beta_2x_{12}^2+\beta_4x_{14}^2+\beta_2\beta_4x_{24}^2\big)^{\de/2-\de_3}}}}
Where the elementary integral over $\beta_3$ was done to get the third
line. A useful integral representation for $I$ is therefore given by
\eqn\Iintrep{I_{\Delta_1\Delta_2\Delta_3\Delta_4}= C \int
d\beta_2d \beta_4
{\beta_2^{\de_2-1}\beta_4^{\de_4-1}\over \big(
x_{13}^2+\beta_2x_{23}^2+\beta_4x_{34}^2\big)^{\de_3}\big
(\beta_2x_{12}^2+\beta_4x\
_{14}^2+\beta_2\beta_4x_{24}^2\big)^{\de/2-\de_3}}}
where the numerical constant $C$ depends on $\de_i,i=1,\cdots,4$.
\eqn\cdef{C= {\pi^2 (\de/2-3)!(\de/2-\de_3-1)!\over 2 (\de_1-1)
!(\de_2-1)!(\de_4-1)!}}
The reprensentation \Iintrep\ will be most useful for the analysis of
the logarithmic singularities involving the four-point functions. The
conformal covariance properties of these intergrals can be made
manifest by expressing one integral in terms of a hypergeometric
functions which depends only on the conformally invariant
cross ratios \cross (see \Muck\FreedmanTwo). 

\listrefs

\end